\magnification=1200
\baselineskip=18truept
\input epsf

\def\preprint{Y}
\def\draftversion{N}

\def\tr{{\rm\ Tr\ }}
\def\chiral{{\bf\rm C}}
\def\wilson{{\bf\rm B}}
\def\ham{{\bf\rm H}}
\def\mbham{{\cal H}}
\def\wblU{{{}^{\rm WB}_{\ \ U}\!\langle}}
\def\wbrU{{\rangle^{\rm WB}_U}}
\def\cap{\hsize=4.5in}
\def\tablerule{\noalign{\hrule}}
\def\htwo{height3pt&\omit && \omit && \omit &\cr}
\def\hthree{height4pt&\omit && \omit && \omit &\cr}

\if \draftversion Y


\fi

\def\figure#1#2#3{\if \preprint Y \midinsert \epsfxsize=#3truein
\centerline{\epsffile{figure_#1_eps}} \halign{##\hfill\quad
&\vtop{\parindent=0pt \hsize=5.5in \strut## \strut}\cr {\bf Figure
#1}&#2 \cr} \endinsert \fi}

\def\captionone{\cap
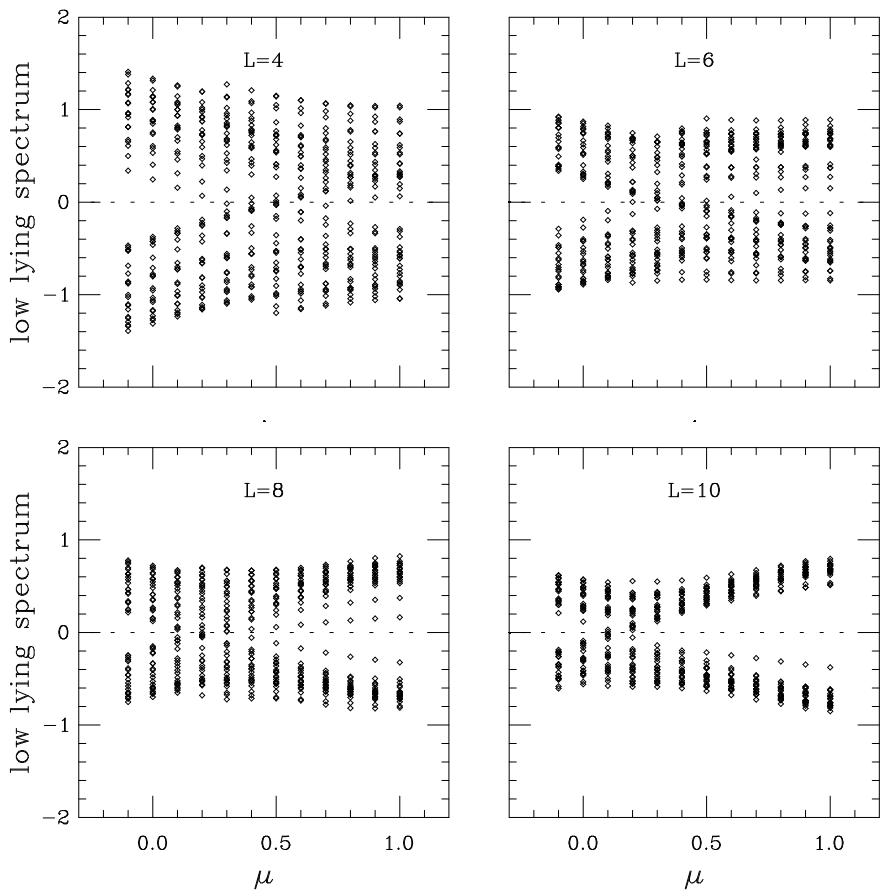
Spectrum of $\ham(\mu)$ as a function of $\mu$ in the two dimensional
U(1) model at a fixed physical volume of ${gL\over\sqrt{\pi}}=3.0$ on four
different $L\times L$ lattices corresponding to four different lattice
spacings proportional to $1\over L$. For each lattice the three low
lying positive and negative eigenvalues of eight different gauge field
configurations are shown.}

\def\captiontwo{\cap
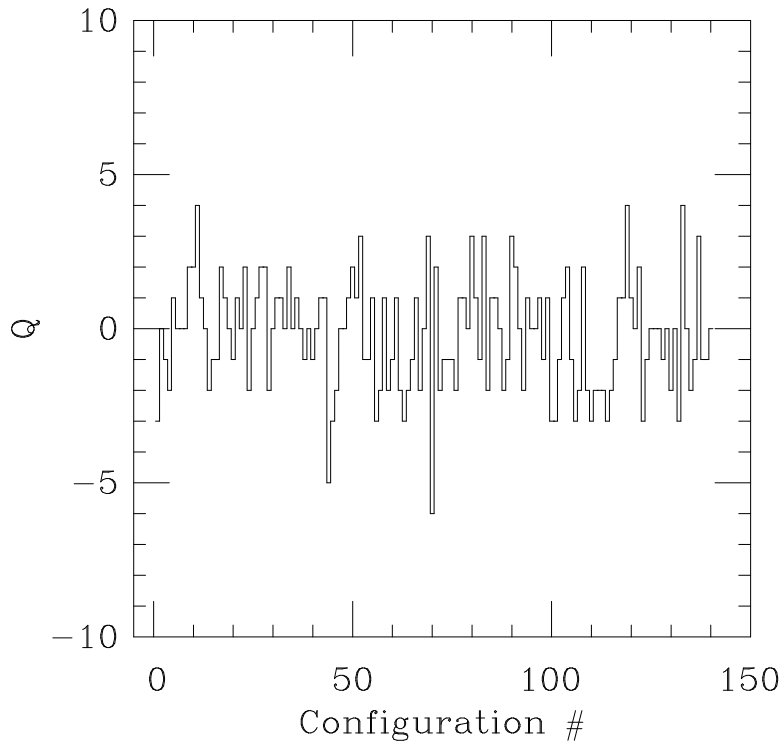
Plot of the index,
$Q(i)$, versus the configuration number, $i$,  as obtained in our
numerical simulation.}

\def\captionthree{\cap 
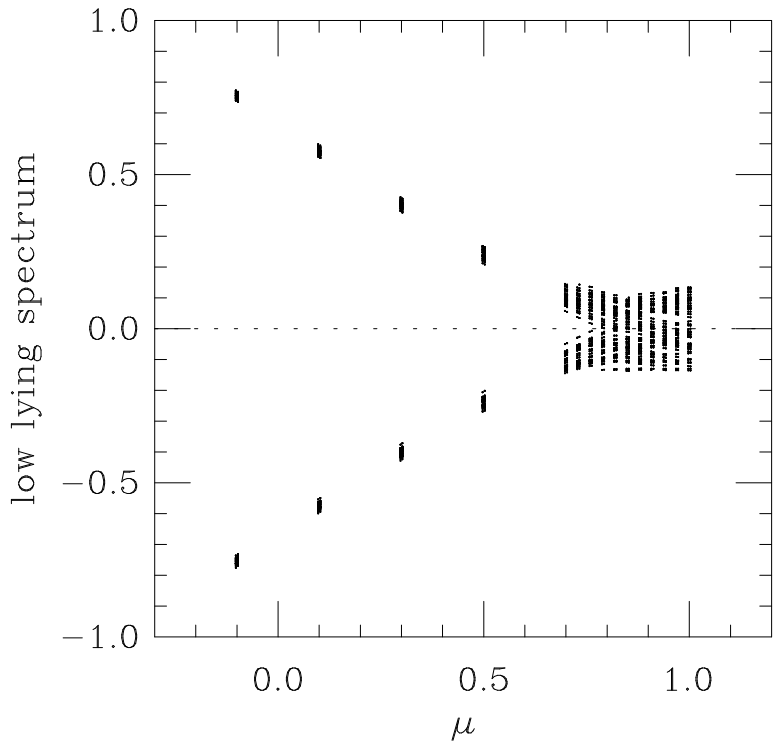
Five low lying positive and negative eigenvalues 
of $\ham(\mu)$ as a function of $\mu$ for five different gauge field
configuration. The configurations are from pure SU(2) gauge theory
at $\beta=2.4$ on a $12^4$ lattice.}

\def\captionfour{\cap 
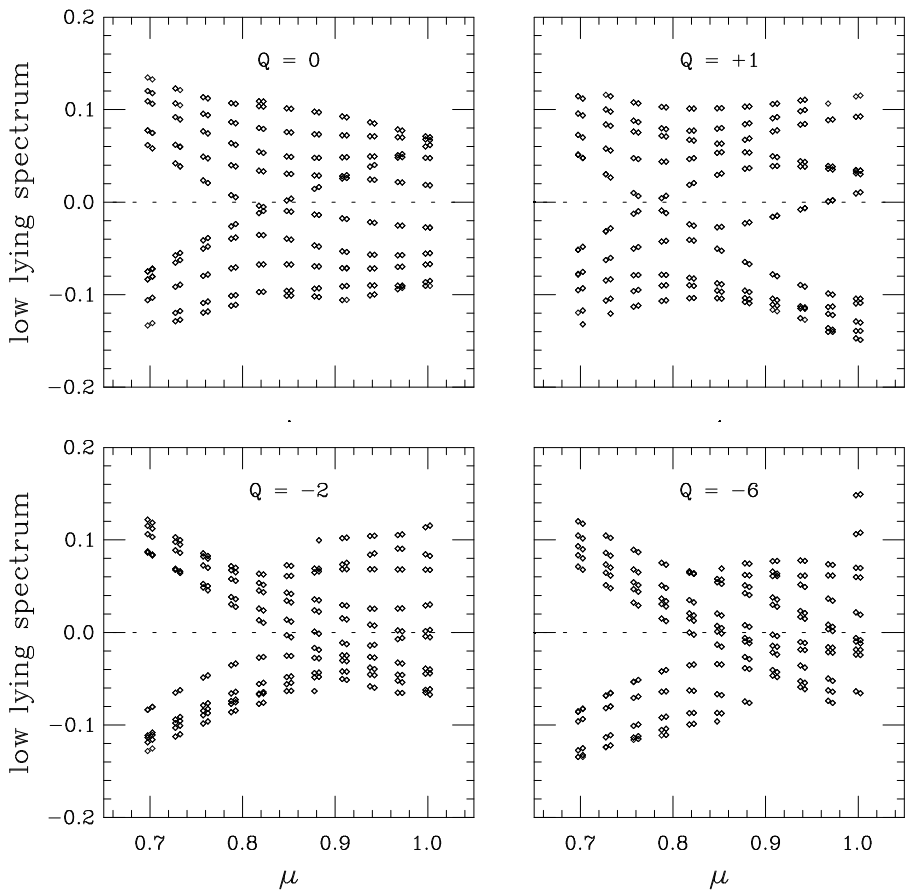
Five low lying positive and negative eigenvalues 
of $\ham(\mu)$ as a function of $\mu$ for four gauge field
configurations with different values of $Q$. The configurations are
from pure SU(2) gauge theory at $\beta=2.4$ on a $12^4$ lattice .  }

\def\captionfive{\cap 
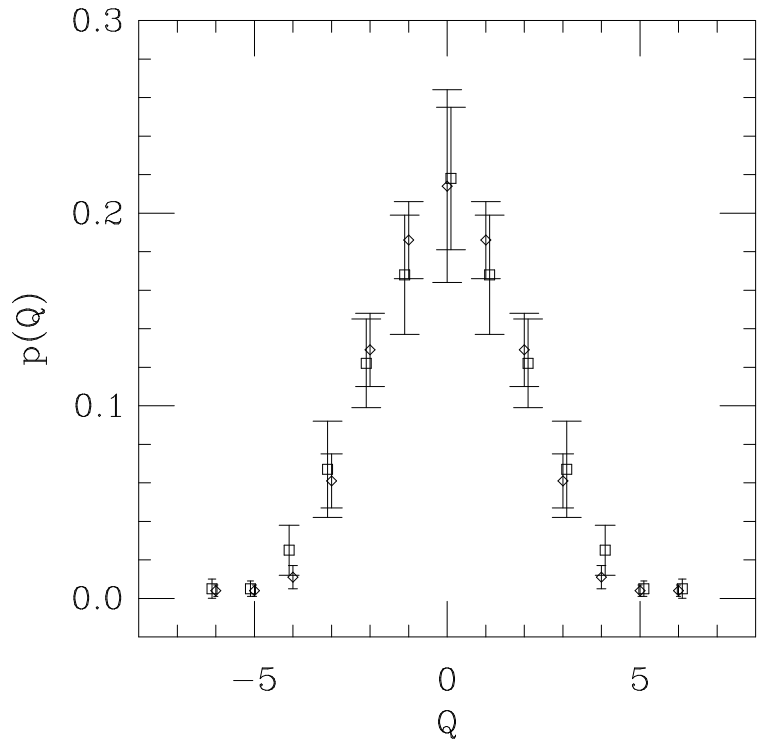
Comparison of the
probability distribution of the index of $\chiral(U)$ obtained here
(diamonds)
with the probability distribution of the topological charge obtained
in [8] (squares). The squares have been slightly shifted laterally
for visual purposes.}

\rightline{UW/PT-97-04}
\rightline{DOE/ER-40561-312-INT97-00-163}
\rightline{CU-TP-811}
\vskip 2truecm
\centerline{\bf A numerical test of the continuum index theorem}
\centerline{\bf on the lattice}

\vskip 1truecm
\centerline{Rajamani Narayanan}
\centerline {\it Institute for Nuclear Theory, Box 351550}
\centerline {\it University of Washington, Seattle, WA 98195-1550}
\bigskip
\centerline{Pavlos Vranas}
\centerline{\it Dept. of Physics, Columbia University}
\centerline{\it New York, NY 10027}

\vskip 1.5truecm

\centerline{\bf Abstract}

The overlap formalism of chiral fermions provides a tool to measure
the index, $Q$, of the chiral Dirac operator in a fixed gauge field
background on the lattice. This enables a numerical measurement of the
probability distribution, $p(Q)$, in Yang-Mills theories.  We
have obtained an estimate for $p(Q)$ in pure SU(2) gauge theory by
measuring $Q$ on $140$ independent gauge field configurations
generated on a $12^4$ lattice using the standard single plaquette
Wilson action at a coupling of $\beta=2.4$. 
This distribution is in
good agreement with a recent measurement [8] of the distribution of
the topological charge on the same lattice using the same coupling and
the same lattice gauge action. In particular we find $\langle
Q^2\rangle=3.3(4)$ to be compared with $\langle Q^2\rangle = 3.9(5)$
found in [8]. The good agreement between the two distributions is an
indication that the continuum index theorem can be carried over in a
probabilistic sense on to the lattice.

\vfill\eject

\centerline{\bf I. Introduction}
\medskip
The four dimensional version of the
Atiyah-Singer index theorem [1] 
states that the index of the
continuum Euclidean chiral Dirac operator,
$$\chiral (A)=\sigma_\mu (\partial_\mu+iA_\mu(x)),\eqno{(1.1)}$$
in a fixed gauge field background, $A_\mu(x)$, is equal to the
topological charge,
$$Q={1\over 32\pi^2} \int d^4x \epsilon_{\mu\nu\rho\sigma}
\tr F_{\mu\nu}(x)F_{\rho\sigma}(x).\eqno{(1.2)}$$
$\sigma_1$, $\sigma_2$ and $\sigma_3$ in (1.1) are the usual
Pauli matrices and $\sigma_4=iI$ where $I$ is the identity matrix.
This theorem has important physical consequences. It provides
an explanation for the baryon number violation in the standard model
[2] and also for the relatively
large value of $\eta^\prime$ in QCD [2,3]. An explicit computation
of baryon number violating processes or of the $\eta^\prime$ mass
needs a non-perturbative technique and the most promising
method is the lattice formalism of gauge theories coupled to
fermions.
An investigation of the index theorem on the lattice is a necessary
precursor to either of the above computations.

In order to investigate the extent to which the index theorem
persists on the lattice, one needs
good definitions for the topological charge of gauge fields
and for the index
of the chiral Dirac operator on the lattice. Unlike the continuum,
there is no division of lattice
gauge fields into disconnected classes since lattice
gauge fields are characterized by group elements on the
links of the lattice and all link variables can
be deformed to unity. Of course, on a lattice there is
no concept of smoothness associated with continuum gauge
fields and therefore the previous statement is not a surprise.
Yet, it is possible to associate an integer with a gauge field
configuration on the lattice in such a way that it matches
with (1.2) for continuum configurations [4]. Instead of using the
geometrically inspired definition on the lattice [4], one could
write down an expression for the integrand in (1.2) on each
site in the lattice and simply sum this up over all lattice sites
to get an expression for $Q$. Such a definition,
referred to as a field theoretic definition, does not result
in an integer value for $Q$ but the probability distribution,
$p(Q)$, is expected to have sharp peaks around integers close
to the continuum limit. Even if this is the case, one expects
difficulties with both the geometric definition and field
theoretic definition due to short distance fluctuations. 
$Q$ measures the topology of the gauge field configuration and is
a globally defined quantity. But in
both the definitions, $Q$ is obtained
as a certain sum over sites on the lattice and
therefore 
large fluctuations of the order of lattice spacing can affect
the measurement of $p(Q)$. This problem has plagued the measurement
of $p(Q)$ for almost a decade.
One approach to resolve this problem,
referred to as the ``cooling method'',  has been to first smooth out
the gauge field configuration and then measure the
topological charge using
the field theoretic method [5]. In principle one could measure the
charge using the geometric definition after cooling but it is not
expected to be different from the field theoretic definition since
the field configuration is smooth. The difficulty with the ``cooling
method'' is the inability to precisely define the stopping criteria
for the cooling procedure. This is due to the fact that the Wilson action
on the lattice has only one minimum corresponding
to zero field density. This difficulty was recently circumvented by
an ``improved cooling algorithm'' [6] based on an improved action [7]
designed to stabilize large instantons on the lattice. Recent
results based on ``improved cooling'' [8] seem to indicate that
this is a good procedure to measure the topological
charge on the lattice. An approach that complements the cooling method
is to measure
directly using the generated
lattice gauge field configurations but improve the field theoretic
operators and also determine the non-perturbative renormalization
of the operator in a systematic manner [9]. Since [8] deals with
SU(2) theory and [9] deals with SU(3) theory a direct comparison
is not possible. Both [8] and [9] deal with the measurement of
the topological charge on the lattice using the 
standard single plaquette Wilson action.
A different approach to circumvent the problem associated with
short distance fluctuations is to improve the action and
measure the topological charge without having to use any
``cooling'' technique [10]. Such a scheme does not seem
to completely remove the problem of
short distance fluctuations and this is resolved in [10]
by interpolating the gauge fields to a finer lattice and using
the geometric method to measure the topological
charge on the fine lattice.
It is possible to compare the result in [8] with the
result in [10] but one has to assume universality and also scaling.
The assumption of universality is needed in the sense that a large class of
actions yield the same continuum limit for
the distribution for the topological charge.
The assumption of scaling is needed since 
only the continuum number in [8] and [10] can be compared.
If one assumes universality and scaling
then the number for the topological susceptibility, $\chi$,
 obtained in [8]
is two to three
standard deviations smaller than the one obtained in [10].

To investigate the remnant of the index theorem on the lattice
and to compare with the different methods used to measure the topological
charge using (1.2) one has to define the index of the chiral
Dirac operator on the lattice. The obvious problem is the
difficulty in formulating chiral fermions on the lattice.
Smit and Vink carefully dealt with the definition of the index 
of $\chiral(A)$ on the lattice [11] by looking at the low lying
eigenvalues of the lattice Dirac operator and the associated
chiralities.
They performed measurements in two dimensional U(1) theory
 using both
Wilson [12] and staggered fermions [13] and in four dimensional
SU(3) theory using staggered fermions [14]. Assuming the
index theorem, the
topological susceptibility, $\chi$, in SU(3) theory was computed 
on the lattice [15] and compared
with the various results for $\chi$ using (1.2).
The comparison revealed that the value for $\chi$ based on the
geometric method [16] was too large and the one obtained from
``cooling'' [5] was not as large but still larger than the one
obtained using the fermionic method [15]. 
The fermionic method 
is inherently global in nature since eigenvalues of
the Dirac operator are
functions of the complete set of gauge fields on the lattice.
In spite of this advantage, there is still a hurdle that one had
to face when using Wilson fermions
or staggered fermions to compute the index of $\chiral(A)$.
This is because one does not have exact zero modes on the
lattice in either case.
Wilson fermions do not have exact zero modes due to the Wilson term
that breaks the chirality and one has to tune the Wilson parameter
per gauge configuration to get a good estimate of the number of
zero modes [12]. The lack of flavor symmetry in staggered fermions
causes a zero mode shift [17] and one has to estimate the
number of zero modes by looking at low lying eigenvalues [13,14].
Owing to this difficulty in estimating the zero modes in a precise
manner, no definite conclusion could be reached from the
comparison in [15]. A comparison of the estimate of $\chi$ in [15]
obtained using the fermionic zero modes is still smaller than the
recent estimate of $\chi$ in [9]. This difference could be due
to improper estimate of zero modes in [15] or due to the
an improper estimate of the non-perturbative renormalization in [9].
On the other hand, if we assume that both [9] and [15] are correct
then the index theorem does not seem to be valid on the lattice
and this would be a serious drawback for the lattice formulation.

The overlap formalism [18] developed to deal with chiral gauge
theories on the lattice provides a natural framework to measure
the index of the chiral Dirac operator as first realized in [19]. 
In this paper
we present the first measurement of the distribution of the index
of the chiral Dirac operator
 in a four dimensional gauge theory using the overlap
formalism. The simplest theory from the
numerical viewpoint is pure SU(2) gauge theory.
Since this is the first measurement of this kind, we decided to
use the standard single plaquette Wilson action and
decided to investigate the continuum
index theorem on the lattice by a direct comparison with one
of the measurements in [8]. 
Therefore we chose a value of $\beta=2.4$
on a $12^4$ lattice. 
This way we do not have to invoke
universality or scaling for the sake of testing the continuum
index theorem on the lattice.
In section II we review the definition of
the index of the chiral Dirac operator using the overlap formalism [18].
The overlap formalism has a parameter $m$ which has to be held fixed
at some value in the range $0<m<1$ as one takes the continuum limit.
In section III, we discuss the role played by this parameter in
defining the index of the chiral Dirac operator. We discuss the
numerical technique and results in section IV.
We measured the index, $Q$, on  $140$ independent configurations and
obtained an estimate for the probability distribution, $p(Q)$. 
This distribution is in
good agreement with a recent measurement [8] of the distribution of
the topological charge on the same lattice using the same coupling and
the same lattice gauge action. In particular we find $\langle
Q^2\rangle=3.3(4)$ to be compared with $\langle Q^2\rangle = 3.9(5)$
found in [8]. The good agreement between the two distributions is an
indication that the continuum index theorem can be carried over in a
probabilistic sense on to the lattice.
We present our conclusions and directions for future
work in section V.

\bigskip
\centerline{\bf II. Index of $\chiral(A)$ in the overlap formalism}
\medskip
The index of the chiral Dirac operator,
$\chiral(A)$, is the difference between the kernel of
$\chiral(A)$ and the kernel of $\chiral^\dagger(A)$. If the
two kernels are not the same then the operator $\chiral(A)$ is
a map between two spaces that differ in dimension. 
Formally, in terms of
the unregulated fermionic action for a single left handed chiral fermion,
$$S_f = \int d^4x \bar\psi_l(x) \chiral(A) \psi_l(x)\eqno{(2.1)}$$
it means that the ``number'' of $\bar\psi_l$ degrees of freedom
is different from the ``number'' of $\psi_l$ degrees of freedom.
Therefore the path integral over all $\bar\psi_l$ and $\psi_l$
yields a zero unless one inserts the excess degrees of freedom
of $\bar\psi_l$ or $\psi_l$ inside the path integral. Since
one has to insert a certain number of $\bar\psi_l$ or $\psi_l$ as
the case may be, one get a non-zero expectation value for a
fermion operator where the operator does not have an equal number
of $\bar\psi_l$ and $\psi_l$. This 
results in the existence of 't Hooft processes [2] in chiral gauge
theories (eg: Weak interactions in the
Standard Model) and vector gauge theories (eg: QCD).

The overlap formalism [18] provides a formula for the generating
functional associated with the path integral of chiral fermions
in a fixed gauge background. It has the important property of
respecting the index of $\chiral(A)$. The overlap formula for the
generating functional
for a left handed chiral fermion coupled to an external gauge field,
$U$, on the lattice is
$$Z(\bar\eta,\eta,U)=\wblU L-|e^{\bar\eta a + \eta a^\dagger}|L+\wbrU
\eqno{(2.2)}$$
where $|L\pm\wbrU$ are many body ground
states\footnote{$^*$}{The superscript, WB, implies a certain phase choice
needed to fully define the states entering in the generating
functional. Since the phase of the overlap does not enter the definition
of the index,
the details of the phase choice will not be of
concern to us in this paper.}
of
$$\mbham^\pm =a^\dagger \ham^\pm(U) a\eqno{(2.3)}$$
$$\ham^\pm(U) = \pmatrix{ \wilson(U)\mp m & \chiral(U) \cr
\chiral^\dagger(U) & -\wilson(U)\pm m \cr }
\eqno{(2.4)}$$
$$
\chiral(x\alpha i, y\beta j;U) =
{1\over 2}
\sum_{\mu=1}^4 \sigma_\mu^{\alpha\beta}
\Bigl [
\delta_{y,x+\hat\mu} (U_\mu(x))^{ij} -
\delta_{x,y+\hat\mu} (U^\dagger_\mu(y))^{ij}
\Bigr] \eqno{(2.5)}
$$
$$
\wilson(x\alpha i, y\beta j;U) =
{1\over 2} \delta_{\alpha\beta}\sum_{\mu=1}^4
\Bigl [
2 \delta_{xy}\delta_{ij} -
\delta_{y,x+\hat\mu} (U_\mu(x))^{ij} -
\delta_{x,y+\hat\mu} (U^\dagger_\mu(y))^{ij}
\Bigr] \eqno{(2.6)}
$$
$0<m<1$ in (2.4) and all values of $m$ in this range
correspond to the same physical theory in the continuum limit.
Different choices of $m$ result in different finite lattice spacing
effects. $\wilson$ is hermitian and so are the Hamiltonians
$\ham^\pm(U)$. 
On a finite lattice, the single particle Hamiltonians, $\ham^\pm(U)$,
are finite matrices and the many body states,
$|L\pm\wbrU$, are obtained by filling all
the negative energy states of  $\ham^\pm(U)$. If the number of
negative energy states of $\ham^-(U)$ are different from
that of $\ham^+(U)$ then the number of bodies making up
$|L-\wbrU$ is different from the number that makes up  $|L+\wbrU$.
We need to differentiate a certain number of times with respect
to $\eta$ or $\bar\eta$ in (2.2) to compensate for this difference
in the number of bodies and get a non-zero result. This difference
in the number of bodies is the index of
$\chiral(U)$.

To understand the connection between the spectrum of $\ham^\pm(U)$
and index of $\chiral(U)$ and also to develop an efficient numerical
algorithm to measure the index using the above connection, it is
useful to consider the eigenvalue flow of
$$\ham(\mu) = \pmatrix{ \wilson -\mu & \chiral \cr 
\chiral^\dagger  & -\wilson +\mu }\eqno{(2.7)}$$
as a function of $\mu$ in a fixed gauge background $U$ (we have
suppressed the dependence on $U$ in (2.7)). Our 
interest is in comparing the spectrum at $\mu=-m$ ($\ham^-=\ham(-m)$) with the 
spectrum at $\mu=m$ ($\ham^+=\ham(m)$). One can prove that $\ham(\mu)$ has
an equal number of positive and negative eigenvalues for all
$\mu < 0$ [18,20]. To see this, we first note that 
$u^\dagger \wilson u \ge 0$  for all vectors $u$
and consider the possibility 
of a zero eigenvalue for $\ham(\mu)$ with eigenvector $\pmatrix {u\cr v\cr}$:
$$\eqalign{  &
(\wilson - \mu) u + \chiral v =0;\ \ \ \ \ \ 
\chiral^\dagger u +(\mu-\wilson) v =0; \ \ \ \ 
u^\dagger u + v^\dagger v = 1 \cr
\Rightarrow & 
u^\dagger \bigl [ (\wilson - \mu) u + \chiral v  \bigr ]
- \Bigl ( v^\dagger \bigl [\chiral^\dagger u +(\mu-\wilson) v 
\bigr ] \Bigr )^* = 0 \cr
\Rightarrow &
u^\dagger \wilson u + v^\dagger \wilson v = \mu \cr}
\eqno{(2.8)}$$
The above equation can have a solution only if $\mu \ge 0$. 
Since the spectrum has an equal number of positive and negative
eigenvalues for $\mu=-\infty$, it follows that the spectrum has
an equal number of positive and negative eigenvalues for
all $\mu < 0$. 

For $\mu \ge 0$, $\ham(\mu)$ can have zero eigenvalues and if this
is the case for $\mu < 1$, then the spectral flow of $\ham(\mu)$
will have some level crossings. Such level crossings
are expected to be of generic type 
in that the first derivative of the crossing
eigenvalue is not zero. Let $n^+$ denote the number of crossings
with positive slope (a negative eigenvalue becoming positive
as $\mu$ is increased) and let $n^-$ denote the number of
crossings with negative slope. If $n^+\ne n^-$, the spectrum
at $\mu=1$ does not have an equal number of positive and negative
eigenvalues implying that the index of $\chiral(U)$ for this
particular $U$ is equal to $Q=n^+-n^-$. 
Given a configuration $U$ with index $Q$ on a finite
lattice, it can be shown that
the parity transformed partner of $U$ has index $-Q$.
This is a consequence of Lemma 4.4 in [18]. This proves
that the distribution of $Q$, namely $p(Q)$ is symmetric
about $Q=0$.

The slope of flow at the crossing from first order
perturbation theory of $\ham(\mu)$ with respect to $\mu$ is
$${d\lambda\over d\mu} = -(u^\dagger u - v^\dagger v)\eqno{(2.9)}$$
where $\pmatrix {u\cr v\cr}$ is the eigenvector with eigenvalue
$\lambda$ at $\mu$. Since
$$u^\dagger u - v^\dagger v = \pmatrix {u^\dagger & v^\dagger\cr}
\gamma_5 \pmatrix {u \cr v \cr}; \ \ \ \
\gamma_5 = \pmatrix {1 & 0 \cr 0 & -1 \cr} \eqno{(2.10)}$$
a positive slope can be associated with one chirality and
a negative slope with the opposite chirality. Such an association
leads to a connection with the continuum definition of the index
of $\chiral(U)$ as was shown in some detail in [18]. 

This shows that one can obtain the index $Q$ of $\chiral(U)$ by
looking at the spectral flow of $\ham(\mu)$, measuring $n^\pm$,
and computing $n^+-n^-$. (2.9) and (2.10) imply that one can associate
a chirality with each crossing. 
Therefore, we can speculate that the gauge field
configuration $U$ is made up of $n^+$ instanton like objects
and $n^-$ anti-instanton like objects based on the level crossings.
If $n^+=n^-$ then
the overlap of the two many body states need not be zero and
hence a non-zero number for $n^+$ in this case
does not necessarily mean the existence of zero modes.

\figure{1}{\captionone}{5.0}

\bigskip
\centerline{ \bf III. The role of $m$}
\medskip
For a given gauge field configuration 
on a finite lattice away from the
continuum, any crossing will occur at some $\mu > 0$
and if there are more than one crossings they will generically
occur for different values of $\mu$. As one goes to the continuum, 
these crossings will occur for smaller values of $\mu$ and approach
$\mu=0$ in the continuum limit indicating a topologically non-trivial
gauge field configuration.  
This can be seen from equation 2.8. Close to the continuum $U_\mu(x)
\sim 1 \Rightarrow B(x,y;U) \ll 1 \Rightarrow u^\dagger B u +
v^\dagger B v = \mu \ll 1$. 
To understand the crossing region in $\mu$ as we approach the
continuum limit,
we study the spectrum as a function of $\mu$ in the 
model with 
$U(1)$ gauge fields on a two dimensional torus [21]. 
The gauge field action
is the standard Wilson plaquette action.
We fix the physical volume, $l^2$,
by chosing a fixed value for ${e l\over \sqrt{\pi}}$ where 
$e$ is the gauge coupling.
To work in a constant physical volume on the lattice,
we impose 
${el\over \sqrt{\pi}}={gL\over \sqrt\pi}$ 
on an $L\times L$ lattice with $g$ being the coupling in lattice units.
The lattice spacing is proportional to $1\over L$.
Fixing 
${gL\over\sqrt{\pi}}=3.0$, we generated several gauge configurations on
$L=4,6,8,10$. In Figure 1, we plot the spectrum of
several configurations where there were level crossings.
As one can readily see, the crossings happen in a small range
of $\mu > 0$ and this range gets closer to
$\mu=0$ as $L$ gets bigger. 

To define the number of crossings and to measure the probability
distribution of the index, $p(Q)$, on a finite lattice we will need to
fix a value of $\mu$ and count only crossings below that $\mu$.  As it
is clear from the above discussion the actual value of $\mu$ that we
pick ($0<\mu<1$) will become unimportant as the continuum limit is
approached since then all crossings will occur in the neighborhood of
$0$. However, the approach of $Q$ toward 
the continuum limit will clearly depend on $\mu$.  For
example, from figure 1 we see that if we set $\mu = 0.4$ then the
$L=4$ case will have almost no crossings below $\mu$ indicating a severe
finite lattice spacing effect.  The $L=6$ case will have some of its
crossings below $\mu$ and the $L=8$, $L=10$ cases will have all of
their crossings below $\mu$. 
On the other hand if we set $\mu\sim 1$ all
four cases have their crossings below $\mu$ and the continuum limit
is approached in a smoother way.  It is therefore advantageous
to choose $\mu \sim 1$ especially since in most cases one is forced to
work far away from the continuum limit.  Since most crossings happen
within a small interval of $\mu$ in practice one scans the full $\mu
\in (0,1)$ interval using a large step size while concentrating in the
region where most crossings occur using a finer step size. For our
study of the SU(2) lattice gauge theory in four dimensions we consider
all crossings below $\mu=1$. 

\bigskip
\centerline{\bf IV. Numerical technique and results}
\medskip
In this section we present results for $p(Q)$ obtained by a direct
measurement of $Q$,
the  index of $\chiral(U)$, using the overlap formalism
in pure SU(2) gauge theory. We use the exponentiated
standard Wilson single
plaquette action as the Boltzmann weight. We have used the
standard heat-bath technique to generate the gauge field configurations [22].
Having generated the gauge field configuration, the index associated
with that configuration is measured by obtaining the flow of $\ham(\mu)$
defined in (2.7). 

\figure{2}{\captiontwo}{5.0}

On an $L^4$ lattice, the hamiltonian is an $8L^4\times 8L^4$ matrix.
Like in the two dimensional U(1) model shown in Figure 1,
the spectrum for the four dimensional
SU(2) theory also has two bands: one is a band of positive eigenvalues
and one is a band of negative eigenvalues. 
Each band has two edges and the spectrum
has four edges. The Lanczos algorithm is designed to efficently extract the
edges of a spectrum and
we used the standard Lanczos algorithm [23] to get the
eigenvalues on these four edges. The part of the spectral flow we
are interested in is associated with 
the two inner edges, namely the two edges close to zero. 
We fix a value of $\mu$ and perform sufficient Lanczos iterations
to obtain  at least five low lying positive and five low lying negative
eigenvalues to a sufficient accuracy. 
We then do the same for several values of $\mu$.  For each value of
$\mu$ we calculate at two points close to each other namely $\mu \pm
0.0025$. This gives the slope of the flow lines at each $\mu$ and
makes the job of identifying the flow lines
considerably easier.
All our crossings happen in the range of $0.7 < \mu < 1.0$ and
so we chose the following set of values for $\mu$: 
$$\eqalign{
\mu=\{ & -0.1,0.0,0.1,0.2,0.3,0.4,0.5,0.6,0.7, \cr
& 0.73,0.76,0.79,0.82,0.85,0.88,
0.91,0.94,0.97,1.0 \}\cr}$$ 
This enables us to perform a careful study of the spectral flow
and estimate the number of $n^+$ and $n^-$ per gauge configuration.

Starting from a random configuration we performed $5000$ heat bath
updates of the lattice to achieve thermalization. We then
measured the index of $140$  gauge field configurations where
two adjacent configurations were separated by $100$ heat bath
updates of the lattice. Having generated a gauge field configuration,
its parity transformed partner has the same action and therefore
a parity transformation will always be accepted by any updating scheme.
We assume that this is done for every configuration and therefore
the distribution, $p(Q)$, is forced to be symmetric. Our data consists
of $140$ configurations and their parity transformed partner. In any
error analysis, we assume that we have only $140$ independent
configurations.

A plot of $Q$ as a function of the
configuration number is shown in figure 2. For this plot we
have ignored configurations obtained by a parity transformation
and is therefore a plot of raw data.
There is no
evidence for any auto-correlation in this plot. In figure 3
we plot an ensemble of the spectral flow that we obtained.
There are a total of five configurations in this ensemble
and depicts the typical spectral flow. The flow is very
similar to the ones obtained in figure 1 for the U(1) model
in two dimensions. In this case, the crossings occur in
the region $0.7 < \mu < 1.0$.
Figure 4 focuses on $0.7\le \mu \le 1.0$
where crossings occur and show four typical configurations
with different values of $Q$. 
For each configuration the number of crossings was deduced by visually
inspecting figures similar to the ones shown in figure 4.  Since our
ensemble contains only $140$ configurations this was not a time
consuming task.  However, for much larger ensembles the process should
be automated.  This will require some form of pattern recognition
and presents an interesting programming problem.

\figure{3}{\captionthree}{5.0}

\figure{4}{\captionfour}{5.0}

{{\null\hfill 
\vbox{\offinterlineskip
\hrule
\halign{\vrule# & #\hfil & \vrule# & #\hfil & \vrule# 
& #\hfil & \vrule# \cr
\hthree
& \ \ $Q$  && \ \ $p(Q)$  && \ \ $p(Q)$[8] & \cr
\hthree
\tablerule\cr
\htwo
& \ \   0 && \ \ 0.214(50) && \ \ 0.218(37) & \cr
\htwo
& \ \   $\pm$1 && \ \ 0.186(20) && \ \ 0.168(31) & \cr
\htwo
& \ \   $\pm$2 && \ \ 0.129(19) && \ \ 0.122(23) & \cr
\htwo
& \ \   $\pm$3 && \ \ 0.061(14) && \ \ 0.067(25) & \cr
\htwo
& \ \   $\pm$4 && \ \ 0.011(6) && \ \ 0.025(13) & \cr
\htwo
& \ \   $\pm$5 && \ \ 0.004(3) && \ \ 0.005(4) & \cr
\htwo
& \ \   $\pm$6 && \ \ 0.004(3) && \ \ 0.005(5) & \cr
\htwo
\hthree
\tablerule\cr
\hthree
& \ \  $\langle Q^2\rangle$ && \ \ 3.3(4) && \ \ 3.9(5) & \cr
\htwo
\hthree}
\hrule}
\hfill}
\vskip .2cm
\item{\bf Table 1:} 
Comparison of the
probability distribution of the index of $\chiral(U)$ obtained here
with the probability distribution of the topological charge obtained
in [8].
\vskip .3cm}

\figure{5}{\captionfive}{5.0}

Since there is data for 
the distribution of topological charge using
improved cooling for the same lattice parameters
($\beta=2.4$ on a $12^4$ lattice) using the same
Wilson single plaquette action a direct comparison
of the two sides of the continuum index theorem
on the lattice is now possible. This comparison is
shown in table 1 and figure 5. The second column in
table 1 is our result for the symmetrized distribution, $p(Q)$
where $Q$ is the index of $\chiral(U)$. The third
column in table 1 is the result from [8] for the
symmetrized distribution, $p(Q)$ where $Q$ is the topological
charge given by (1.2). The two columns are a result
of measurements on a different set of independent
configurations. 
The close matching of the two distributions indicate
that the connection between the index of $\chiral(U)$ and
the topological charge of $Q(U)$ remains valid on the
lattice in a probabilistic sense. 
The variance of $Q$, namely $\langle Q^2 \rangle$, obtained
here and in [8] is also listed in Table 1
and they agree quite well. 
If we use a lattice spacing of $a=0.12$fm associated with
$\beta=2.4$, then we get $\chi=184(6)$MeV. 
To properly verify this as a continuum result we will have
to show evidence for scaling in $p(Q)$ where $Q$ is the
index of $\chiral(U)$.

Now we turn our attention to $p(n^+,n^-)$, the probabilty of
finding a configuration with $n^+$ instantons like objects
and $n^-$ anti-instanton like objects. Such a configuration
is expected to have $(n^++n^-)$ localized objects 
and the Dirac operator is expected to have
the same number of localized
eigenmodes. We assume that every
crossing in a typical flow like the ones in figure 4 is associated
with a localized eigenmode of the Dirac operator. By looking at the
number of crossings with positive slope and number of crossings with
negative slope, we obtain the $n^+$ and $n^-$ associated with that
gauge configuration. Given a configuration with a fixed $n^+$ and
$n^-$ we know that the parity transformed gauge field will have
$n^+$ and $n^-$ interchanged. Therefore $p(n^+,n^-)=p(n^-,n^+)$
and we have listed $p(n^+,n^-)$ for $n^+ \le  n^-$ in Table 2
as obtained from the $140$ independent measurements. 
In our sample we did not find any configuration with 
more than six objects. From the 
distribution we obtain $\langle (n^+ + n^-) \rangle = 2.3(1)$
implying that on an average the gauge field configuration has roughly
two to three localized objects. 
We have obtained this number by
simply taking the gauge field configuration without applying
any cooling or smoothing algorithm. A similar result has also
been obtained in [8] but their result is a function of the cooling
sweeps. It is interesting to note that our result for  
$\langle (n^+ + n^-) \rangle$ is compatible with the result 
obtained in [8] after $150$ cooling sweeps. It would be interesting to
obtain the scaling form of $p(n^+,n^-)$ using the method described here
and compare it with a similar result using the method in [8]. A natural
assumption is that the result for $p(n^+,n^-)$ in [8] would be
unaffected by cooling if the lattice coupling is well within the
scaling region and would be in agreement with the result obtained using
the overlap formalism.
$p(n^+,n^-)$ has also information about interactions between instantons
and anti-instantons. Let us assume that all instantons are identical
and all anti-instantons are identical. A model of non-interacting
objects is consistent with our data.
For example, we see that $2 p(0,2)=p(1,1)$ within
errors indicating no difference in interactions between like 
and unlike objects. 
Also we see that $2 p^2(0,2) = 3 p(0,1) p(0,3)$ 
within errors
indicating that there is no interaction between like objects. 
Of course
the errors in the data in table 2 are too large to make any definite
conclusion or help in building an emperical model for the interaction
of instantons and anti-instantons. 
We see that the zero topological sector is dominated
by configurations with $n^+\ne 0$ and this is thought be relevant for
a study of chiral symmetry breaking [24].
A better estimate of $p(n^+,n^-)$ would facilitate a comparison with
existing models for instanton--anti-instanton interactions 
[24].

{\null\hfill 
\vbox{\offinterlineskip
\hrule
\halign{\vrule# & #\hfil & \vrule# & #\hfil & \vrule# 
& #\hfil & \vrule# \cr
\hthree
& \ \ $n^+$  && \ \ $n^-$  && \ \ $p(n^+,n^-)$ & \cr
\hthree
\tablerule\cr
\htwo
& \ \   0 && \ \ 0 && \ \ 0.036(16) & \cr
\htwo
& \ \   0 && \ \ 1 && \ \ 0.114(15) & \cr
\htwo
& \ \   0 && \ \ 2 && \ \ 0.104(19) & \cr
\htwo
& \ \   0 && \ \ 3 && \ \ 0.057(13) & \cr
\htwo
& \ \   0 && \ \ 4 && \ \ 0.007(5) & \cr
\htwo
& \ \   0 && \ \ 5 && \ \ 0.004(3)  & \cr
\htwo
& \ \   0 && \ \ 6 && \ \ 0.004(3)  & \cr
\htwo
& \ \   1 && \ \ 1 && \ \ 0.164(39) & \cr
\htwo
& \ \   1 && \ \ 2 && \ \ 0.057(12) & \cr
\htwo
& \ \   1 && \ \ 3 && \ \ 0.025(12) & \cr
\htwo
& \ \   1 && \ \ 4 && \ \ 0.004(3)  & \cr
\htwo
& \ \   1 && \ \ 5 && \ \ 0.004(3)  & \cr
\htwo
& \ \   2 && \ \ 2 && \ \ 0.014(10) & \cr
\htwo
& \ \   2 && \ \ 3 && \ \ 0.014(6) & \cr
\htwo
\hthree
\tablerule\cr
}
\hrule}
\hfill}
\vskip .2cm
\item{\bf Table 2:} 
Probability distribution of configurations with $n^+$ instanton like
objects and $n^-$ anti-instanton like objects. This distribution
yields $\langle (n^+ + n^-) \rangle =2.3(1)$. 
\vskip .3cm

\bigskip
\centerline{\bf V. Conclusions}
\medskip
In this paper we have used the overlap formalism [18] developed
to deal with determinants of chiral Dirac operators as a tool to
measure the Atiyah-Singer index of lattice gauge fields. By working
with the standard single plaquette action on a $12^4$ lattice
with a coupling of $\beta=2.4$ we could directly compare the
probability distribution of $Q$, the index of $\chiral(U)$, obtained
here with the probability distribution of the topological charge
obtained in [8]. Table 1 provides this comparison.
The good agreement between the two distributions
is an indication that the continuum index theorem can be carried over
in a probabilistic sense on to the lattice. 

In this paper we have only done a single coupling on a single lattice.
More such measurements have to be done to establish scaling of the
probability distribution of $Q$
and follow the index theorem to the continuum. 
The algorithm to measure the index
involves a Lanczos procedure and the Hamiltonian operators involved
are directly related to the usual Wilson-Dirac operators. Our code
was written for a vector machine. Each measurement of $Q$ on a single
gauge configuration involved $2000$ Lanczos iterations and took a
total of 2.75 hours of CPU time on a single processor of
the CRAY C90 machine. The time will scale 
linearly with the number of lattice points and linearly with
the number of Lanczos iterations. It is possible to study scaling
in a systematic manner with the use of present day computers.
This work is currently in progress.

\noindent {\bf Acknowledgments}:
The research of R. N. was supported in part by the
DOE under grant \# DE-FG03-96ER40956 and \# DE-FG06-90ER40561 and that of
P. V. was supported in part by the DOE under grant \#
DE-FG02-92ER40699. 
This research was also supported in part by grant number PHY960006P 
from the Pittsburgh Supercomputing Center, 
sponsored by the National Science Foundation (NSF).
We would like to thank Margarita Garc\'{\i}a P\'erez for useful discussions
and providing us with data
for Table 1. We would also like to thank
Herbert Neuberger, Philippe de Forcrand and Ion-Olimpiu Stamatescu
for useful discussions.

\if \preprint Y
\vskip .3in
\fi
\if \preprint N
\vskip 1in
\fi
\noindent{\bf References}
\vskip .2in
\item{[1]} M. Atiyah, I. Singer, Ann. Math. 87 (1968) 484.
\item{[2]} G. 't Hooft, Phys. Rev. Lett 37 (1976) 8;
Phys. Rev. D14 (1976) 3432; Phys. Reports 142 (1986) 357.
\item{[3]} E. Witten, Nucl. Phys. B156 (1979) 269;
G. Veneziano, Nucl. Phys. B159 (1979) 213.
\item{[4]} M. L\"uscher, Comm. Math. Phys. 85(1982) 39.
\item{[5]} B. Berg, Phys. Lett. B104 (1981) 475;
J. Hoek, M. Teper, J. Waterhouse, Phys. Lett. B180 (1986) 112;
J. Hoek, M. Teper, J. Waterhouse, Nucl. Phys. B288 (1987) 589;
M. Teper, Phys. Lett. B202 (1988) 553.
\item{[6]} 
Ph. de Forcrand, M. Garc\'{\i}a P\'erez and I.-O. Stamatescu, 
hep-lat/9608032.
\item{[7]} M. Garc\'{\i}a P\'erez, A. Gonzalez-Arroyo, J. Snippe, and 
P. van Baal,
Nucl. Phys. B413 (1994) 535.
\item{[8]} Ph. de Forcrand, M. Garc\'{\i}a P\'erez and I.-O. Stamatescu, 
hep-lat/97011012. 
\item{[9]} B. Alles, G. Boyd, M. D'Elia, A. Di Giacomo, hep-lat/9610009.
\item{[10]}
T. DeGrand, A. Hasenfratz, D. Zhu, hep-lat/9607082;
T. DeGrand, A. Hasenfratz, D. Zhu, hep-lat/9604018.
\item{[11]} J. Smit, J. Vink, Nucl. Phys. B286 (1987) 485.
\item{[12]} J. Vink, Nucl. Phys. B307 (1988) 549.
\item{[13]} J. Smit, J. Vink, Nucl. Phys. B303 (1988) 36.
\item{[14]} J. Smit, J. Vink, Phys. Lett. B194 (1987) 433.
\item{[15]} J. Vink, Phys. Lett. B212 (1988) 483.
\item{[16]} M. G\"ockeler, A.S. Kronfeld, M.L. Laursen, G. Schierholz,
U.-J. Wiese, Nucl. Phys. B292 (1987) 349;
A.S. Kronfeld, Nucl. Phys. B (Proc. Suppl.) 4 (1988) 329;
M. G\"ockeler, A.S. Kronfeld, M.L. Laursen, G. Schierholz,
U.-J. Wiese, Phys. Lett. B209 (1988) 315.
\item{[17]} J. Vink, Phys. Lett. B210 (1988) 211.
\item{[18]} R. Narayanan, H. Neuberger, Nucl. Phys. B443 (1995) 305.
\item{[19]} R. Narayanan, H. Neuberger, Nucl. Phys. B412 (1994) 574.
\item{[20]} R. Narayanan, H. Neuberger, Phys. Rev. Lett. 71 (1993) 3251.
\item{[21]}  R. Narayanan, H. Neuberger, P. Vranas, 
Phys. Lett. B353 (1995) 507;  R. Narayanan, H. Neuberger, P. Vranas, 
Nucl. Phys. B (Proc. Suppl.) 47 (1996) 596.
\item{[22]} M. Creutz, {\sl Quarks, Gluons and Lattices}, Cambridge
University Press, 1983.
\item{[23]} G.H. Golub and C.F. Van Loan, {\sl Matrix computations},
Johns Hopkins University Press, 1989.
\item{[24]} For a recent review, see 
T. Schafer, E.V. Shuryak, hep-ph/9610451 and references therein.
\if \preprint N
\vskip 1in
\noindent{\bf Figure Captions}

\halign{#\hfill\qquad &\vtop{\parindent=0pt \hsize=5in \strut#
\strut}\cr
Figure 1: & \captionone\cr
&\cr}

\vskip .5in

\figureb{1}{7.0}

\fi
\vfill\eject
\end